\documentclass[aps,prl,showpacs,preprintnumbers,amsmath,amssymb,twocolumn,letterpaper,superscriptaddress]{revtex4-1}
\usepackage{amssymb}
\usepackage{ifpdf}
\usepackage[utf8]{inputenc}
\usepackage{graphicx}
\usepackage{times}
\usepackage{dcolumn}
\usepackage{float}
\usepackage{bibunits}

\makeatletter
\newcommand*{\balancecolsandclearpage}{%
  \close@column@grid
  \clearpage
  \twocolumngrid
}
\makeatother

\def\CeRh{CeRhIn$_{5}$}
\def\CeCo{CeCoIn$_{5}$}

\begin{document}
\title{The magnitude of the magnetic exchange interaction in the heavy fermion antiferromagnet \CeRh}

\author{Pinaki Das}
\affiliation{Condensed Matter and Magnet Science, Los Alamos National Laboratory, Los Alamos, New Mexico 87545, USA}
\author{S.-Z. Lin}
\affiliation{T-4, Los Alamos National Laboratory, Los Alamos, New Mexico 87545, USA}
\author{N. J. Ghimire}
\affiliation{Condensed Matter and Magnet Science, Los Alamos National Laboratory, Los Alamos, New Mexico 87545, USA}
\author{K. Huang}
\affiliation{Condensed Matter and Magnet Science, Los Alamos National Laboratory, Los Alamos, New Mexico 87545, USA}
\affiliation{Department of Physics, University of California, San Diego, La Jolla, California 92093, USA}
\author{F. Ronning}
\author{E. D. Bauer}
\author{J. D. Thompson}
\affiliation{Condensed Matter and Magnet Science, Los Alamos National Laboratory, Los Alamos, New Mexico 87545, USA}
\author{C. D. Batista}
\affiliation{T-4, Los Alamos National Laboratory, Los Alamos, New Mexico 87545, USA}
\author{G. Ehlers}
\affiliation{Quantum Condensed Matter Division, Oak Ridge National Laboratory, Oak Ridge, Tennessee 37831-6475, USA}
\author{M. Janoschek}
%\email[Corresponding Author: ]{mjanoschek@lanl.gov}
\affiliation{Condensed Matter and Magnet Science, Los Alamos National Laboratory, Los Alamos, New Mexico 87545, USA}

\date{\today}

\begin{abstract}
We have used high-resolution neutron spectroscopy experiments to determine the complete spin wave spectrum of the heavy fermion antiferromagnet \CeRh. The spin wave dispersion can be quantitatively reproduced with a simple $J_1$-$J_2$ model that also naturally explains the magnetic spin-spiral ground state of \CeRh\, and yields a dominant in-plane nearest-neighbor magnetic exchange constant $J_0$~=~0.74 meV. Our results pave the way to a quantitative understanding of the rich low-temperature phase diagram of the prominent Ce$T$In$_5$ ($T$ = Co, Rh, Ir)  class of heavy fermion materials.
\end{abstract}

%abstract length limit: 600 characters

% PACS, the Physics and Astronomy Classification Scheme.
%71.27.+a: Strongly correlated electron systems; heavy fermions
%74.70.Tx: Heavy-fermion superconductors (for heavy-fermion systems in magnetically ordered materials, see 75.30.Mb: see also 71.27.+a Strongly correlated electron systems, heavy fermions)
%75.30.Mb: Valence fluctuation, Kondo lattice, and heavy-fermion phenomena (see also 71.27.+a Strongly correlated electron systems, heavy fermions; for heavy-fermion superconductors, see 74.70.Tx)
%75.30.Ds spin waves,
%75.30.Et Exchange interactions: magnetically ordered materials, 75.30.Et

\vskip2pc

\maketitle
\begin{bibunit}
The strength of magnetic interactions in heavy fermion (HF) materials represents a key parameter that controls the delicate interplay of localized and itinerant electronic degrees of freedom, which drives a multitude of fascinating strongly correlated electron phenomena in these materials \cite{Loehneysen07, Gegenwart08}. This competition between localized $f$ electrons of Ce, Pr, Yb or various actinide elements on one hand and the conduction electrons on the other, is qualitatively described by the Doniach phase diagram \cite{Doniach77}: On one end the magnetic Ruderman-Kittel-Kasayu-Yosida (RKKY) interaction couples neighboring local magnetic moments associated with the $f$-electrons via the conduction electrons and leads to the development of long-range antiferromagnetic (AFM) order. In the other limit, the Kondo interaction~\cite{Kondo64} drives the local demagnetization of the $f$-electron state that is quenched by the spins of the surrounding conduction electrons. This results in delocalization of the $f$-electrons into the conduction band and thus the formation of paramagnetic heavy Fermi liquid state \cite{Fisk86}.

Frequently the most interesting situation occurs near the boundary between the AFM and the demagnetized heavy Fermi liquid state at which Kondo and RKKY interactions mutually cancel~\cite{Doniach77}. At this point the ordering temperature is suppressed to zero resulting in a quantum critical point (QCP) at which the associated long-wavelength quantum critical magnetic fluctuations are believed to lead to the emergence of novel states of matter~\cite{Loehneysen07, Gegenwart08}, with the most prominent one being unconventional superconductivity \cite{Mathur98, Pfleiderer09}. This mechanism is also relevant for other classes of unconventional superconductors such as the cuprates or iron-pnictides \cite{Dahm09, Wang13}.

Apart from this prototypical QCP scenario in which the AFM order parameter becomes critical, more recently so-called local quantum criticality has been invoked theoretically. In the latter, the local Kondo effect becomes critical at a QCP resulting in local instead of long-wavelength magnetic fluctuations~\cite{Si01}. It is further proposed that the position of the QCP may be controlled via the amount of magnetic frustration in a material \cite{Coleman10,Si14}. Because HF materials are usually not geometrically frustrated the magnetic frustration would likely originate from a frustration between nearest and further-neighbor magnetic interactions, highlighting that quantitative knowledge of the magnetic interactions is critical for HF compounds.

Despite the key role of magnetic interactions for the understanding of HF physics, so far little attention has been devoted to determine them quantitatively in HF materials, and few neutron spectroscopy studies have been carried out so far~\cite{Broholm87,Van Dijk00, Knafo03,Fak08,Stockert11}. Here we report an extensive neutron spectroscopy study carried out on the HF antiferromagnet \CeRh\ that reveals the strength of the RKKY interaction in this system. \CeRh\ belongs to the intensively studied family of HF materials Ce$T$In$_5$ ($T$ = Co, Rh, Ir) for which external pressure $P$, magnetic field $H$, and chemical substitution have been demonstrated to tune the ratio between RKKY and Kondo interactions, in turn allowing to access multiple AFM and superconducting phases, as well as various QCPs \cite{Thompson:12}.

While our study has important implications for the understanding of HF materials in general, its aim is to shed more light on the complex interplay between antiferromagnetism,  unconventional superconductivity and quantum criticality in \CeRh\ by probing the strength of the magnetic exchange interactions. Below $T_N$ = 3.8 K, \CeRh\ exhibits incommensurate AFM order with a propagation vector $\mathbf{k}$~=~($0.5~0.5~0.297$) \cite{Bao00}. The AFM order can be suppressed via the application of pressure leading to a QCP at $P_{\rm{c}}$ = 23 kbar around which a broad superconducting dome emerges with a maximal $T_c$ = 2.3 K~\cite{Park06,Park08}. While the recent report of a spin resonance in the superconducting state of \CeCo\ illustrates that magnetic fluctuations are relevant for the superconductivity observed in the Ce$T$In$_5$ family, transport~\cite{Park08} as well as de Haas-van Alphen measurements ~\cite{Shishido05} suggest that other critical degrees of freedom such as charge fluctuations may also be present at the QCP.

\begin{figure}[th]
 \centering
 \includegraphics[width=0.35\textwidth]{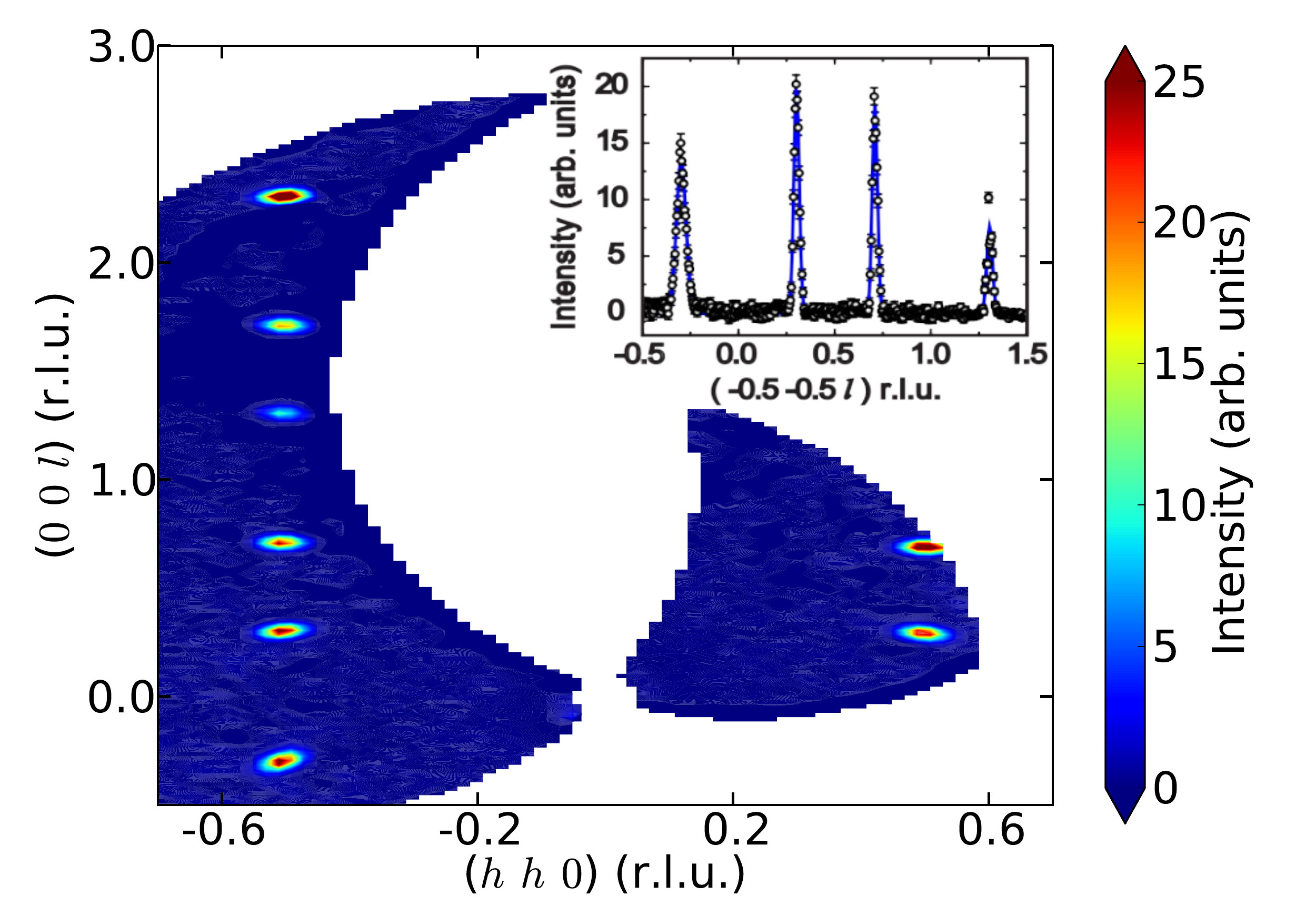}
 \caption{A map of the magnetic elastic scattering in \CeRh\ at a temperature $T$~=~2.5 K. Reference data obtained at $T$~=~20~K was subtracted from the data (see text). A cut along ($-0.5~ -0.5~ l$) shows clear satellite peaks that can be indexed with the propagation vector $\mathbf{k}$~=~($0.5~0.5~0.297$) (inset).}\label{fig:elastic}
\end{figure}

To determine the size of the magnetic exchange interaction we have measured the complete spin wave spectrum of \CeRh\ using the Cold Neutron Chopper Spectrometer (CNCS) at the Spallation Neutron Source \cite{Ehlers11}. To overcome the issues related to the high neutron absorption of both Rh and In, large single crystals of \CeRh\ of typical sizes of 10 x 10 x 5 mm$^3$ were grown via the In self-flux method, and cut and polished to an average thickness of 0.6 mm along the crystallographic $c$ axis to optimize the neutron transmission for cold neutrons. The intensity was further maximized by co-aligning 14 crystals that were fixed on a thin aluminum plate to cover the entire incident neutron beam cross-section of 50 $\times$ 15 mm$^2$ at the sample position, resulting in a total sample mass of 2.2 g. The mosaic spread of the sample was verified by x-ray diffraction and was found to be $1.5^{\circ}$ illustrating the excellent quality of the mosaic assembly. The sample was aligned with the (110) and (001) reciprocal lattice directions in the scattering plane, which in combination with the detector tubes at CNCS that cover a vertical range of $\pm$15$^{\circ}$ perpendicular to the scattering plane, enabled us to probe the spin wave dispersion along the three main high-symmentry directions ($h$00), ($hh$0) and (00$l$). All measurements reported were carried out at a temperature of 2.5 K and with an incident energy $E_i ~=~ 3.315$ meV providing an energy resolution $\Delta E$~$\approx$~80~$\mu$eV at the elastic line. From all plots shown here a reference data set recorded at $T$ ~=~ 20 K was subtracted to highlight the scattering from magnetic correlations. $T$ ~=~ 20 K was chosen because the quasielastic magnetic fluctuations in \CeRh\ extend to about 3$T_N$ \cite{Bao02}.

Fig.~\ref{fig:elastic} shows a map of the magnetic elastic scattering as observed in the scattering plane with clear satellite peaks at $\mathbf{\tau}\pm\mathbf{k}$ (where $\mathbf{\tau}$ is a reciprocal lattice vector) that have been previously shown to correspond to an AFM spin-spiral propagating along the crystallographic $c$-axis~\cite{Bao00}. The inset shows a cut along the ($-0.5~ -0.5~ l$) direction revealing sharp Bragg peaks with an average FWHM of 0.025 \AA$^{-1}$ confirming the excellent quality of the assembled sample mosaic. We also note that the background is flat and zero demonstrating that subtracting the $T$ ~=~ 20 K reference data works reliably. In order to maximize the neutron intensity, all inelastic measurements were carried out around the magnetic zone center ($-0.5~ -0.5~ 0.297$), where the magnetic form factor for Ce$^{3+}$ is large. This additionally allowed us to keep the angle $\theta$ between the crystallographic $c$-axis and the incident beam as low as possible to limit neutron absorbtion due to the increased path length of neutrons traversing the samples for higher $\theta$.

\begin{figure*}[th]
  \centering
  \includegraphics[width=0.85\textwidth]{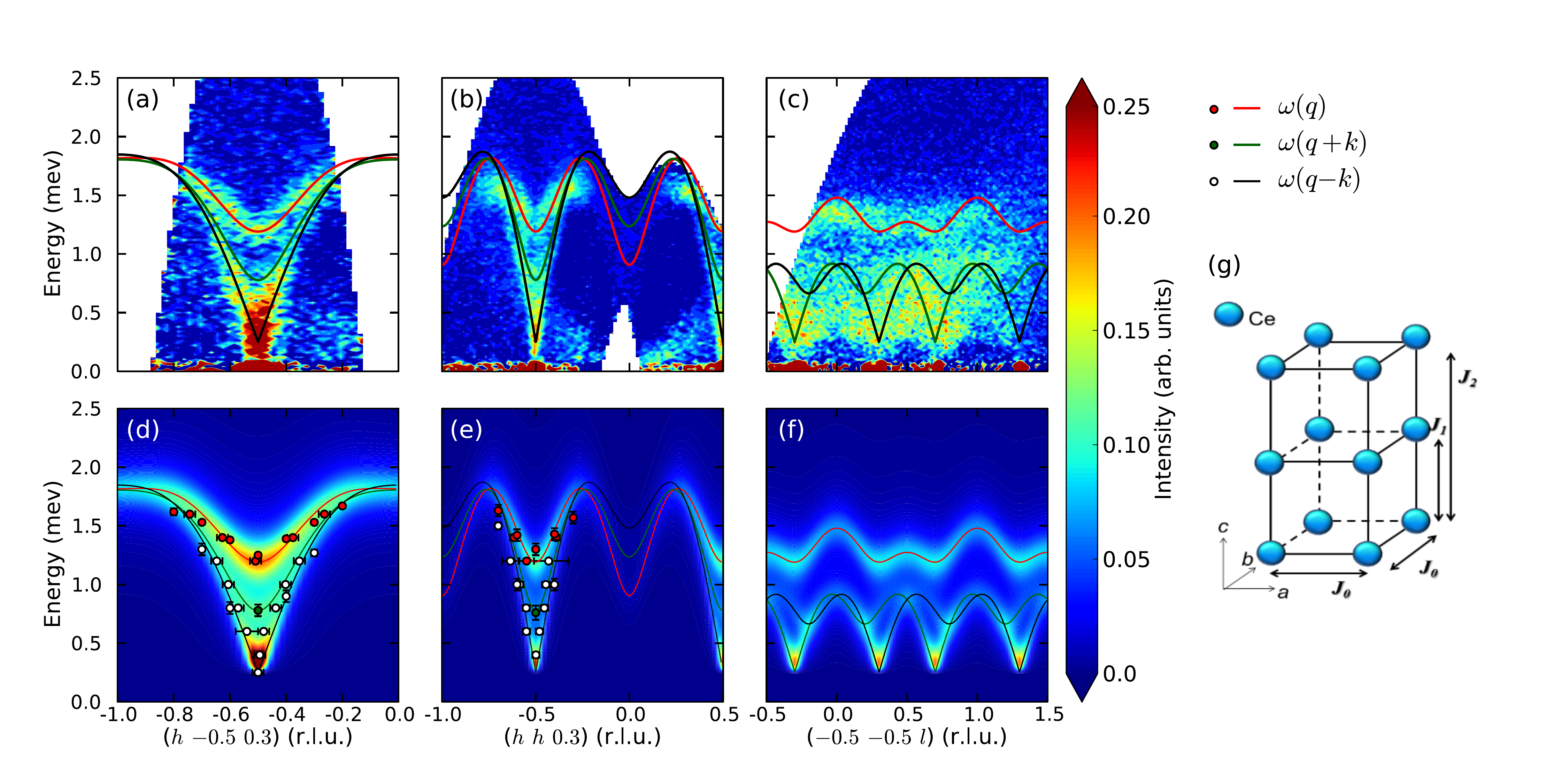}
  \caption{The measured spin wave spectrum compared to calculations based on a simple $J_1$-$J_2$ model that also describes the antiferromagnetic spin-spiral ground state of \CeRh (see text). Panels (a)-(c) show the experimental spin wave spectrum obtained at a temperature $T$~=~2.5 K for the three principal directions ($h00$), ($hh$0) and (00$l$), respectively. $T$~=~20 K reference data was subtracted from each set (see text). The red, green and black solid lines and symbols denote the three branches $\omega_0(\mathbf{q})$, $\omega_0(\mathbf{q}~ +~ \mathbf{k})$, and $\omega_0(\mathbf{q}~ -~ \mathbf{k})$ of the dispersion corresponding to our spin wave model (see text for details) and have been fitted to our data. Panels (d)-(f) show calculations of the full magnetic susceptibility for the same directions as the data presented in (a)-(c) (see text). The circles correspond to the spin wave peak positions extracted from momentum and energy transfer scans as the ones shown in Fig.~\ref{fig:cuts}. In (g) we denote the three magnetic exchange constants used for our calculation.}\label{fig:spinwave}
\end{figure*}

Panels (a)-(c) of Fig.~\ref{fig:spinwave} illustrate the full spin wave spectrum observed in \CeRh\ as measured along the three principal directions ($h00$), ($hh$0) and (00$l$), respectively. The salient features that can be derived from the data without theoretical assumptions are: (i) The bandwidth of the spin wave spectrum is approximately 1.8 meV. While the data shown here only cover energy transfers up to 2.5 meV, we have performed additional measurements with $E_i$~=~12 meV that demonstrate the absence of higher spin wave branches to at least 10 meV. (ii) The spin wave spectrum exhibits at least two distinct branches for the ($h00$) and ($hh$0) directions. (iii) There is a small spin wave gap $\Delta$ of about 0.25 meV.

\begin{figure}[ht!]
 \centering
  \includegraphics[width=0.45\textwidth]{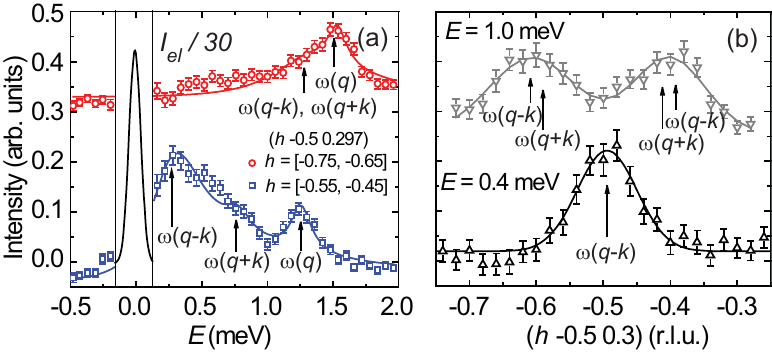}
  \caption{Constant momentum (a) and energy transfer (b) cuts for the for the ($h$00) direction. The intensity of the elastic line in (a) is scaled by a factor of 30. The different scans are shifted vertically for clarity. The arrows denote the various spin wave branches extracted from fits (solid lines, see text for details).}\label{fig:cuts}
\end{figure}

In order to investigate the observed spin wave in detail we show constant momentum and energy transfer cuts for the ($h$00) direction in Figs.~\ref{fig:cuts}(a) and (b), respectively. For the energy scans in Fig.~\ref{fig:cuts}(a) the spin wave peaks were fitted using a Lorentzian line shape. The scan at the zone center (blue squares) clearly reveals a small spin wave gap $\Delta$~=~0.25(3) meV. Further, as indicated by the three arrows, there are actually not two but three distinct spin wave branches, which is in agreement with a model for the spin waves as demonstrated below. For higher momentum transfers (red circles in Fig.~\ref{fig:cuts}(a))~the two lower branches overlap and cannot be resolved with the resolution of this experiment. The peaks in the momentum transfer scans in Fig.~\ref{fig:cuts}(b) were fit with a Gaussian line shape. Similar cuts were also performed for the ($hh$0) symmetry direction, and the peak positions extracted from all cuts are shown as symbols in panels (d)-(e) of Fig.~\ref{fig:spinwave} overlayed on the theoretical calculation of the full spin wave spectrum. We note that for the (00$l$) direction the resolution available at CNCS is not sufficient to perform such fits.

We now turn to the calculation of the spin wave spectrum illustrated in Fig.~\ref{fig:spinwave}. The interaction between the localized Ce magnetic moments is mediated by the conduction electrons, and therefore is described by an RKKY exchange interaction $J(\mathbf{q})$. As illustrated in Fig.~\ref{fig:elastic}, $J(\mathbf{q})$ is maximal when the reduced wave vector $\mathbf{q}$~=~$\mathbf{k}$~=~($0.5~0.5~0.297$). The associated spin spiral ground state configuration can be reproduced with the following minimal $J_1$-$J_2$ model, for which we employ a total of three magnetic exchange constants as illustrated in Fig.~\ref{fig:spinwave}(g): $J_0>0$ is the intralayer nearest-neighbor (NN) antiferromagnetic interaction, $J_1$ is the interlayer NN interaction, and $J_2$ is the interlayer next NN interaction. We therefore obtain the following model Hamiltonian
\begin{equation}
\label{hamil}
   \mathcal{H}=\sum _{{i<j}}J_{{ij}}\left[ \left(S_{x, i} S_{x, j} + S_{y, i}S_{y, j}\right)+\delta S_{z, i}S_{z, j} \right],
\end{equation}
with $J(\mathbf{q})=2 J_0 [\cos(2\pi q_x)+\cos  (2\pi q_y)]+2 J_1 \cos (2\pi q_z)+2 J_2 \cos  (4\pi q_z)$. Here $\mathbf{S}_i$ is a spin 1/2 operator  representing the effective magnetic moment of the $\Gamma^2_7$ doublet. Because the energy gap to the first excited doublet ($\Delta_{\Gamma^1_7} \simeq 7$meV) is much bigger than the exchange constants $J_{ij}$~\cite{Willers10}, we can restrict our model to the lowest-energy doublet. Because the observed  magnetic moments rotate in the tetragonal basal plane of~\CeRh~\cite{Bao00}, we assume an  easy-plane exchange anisotropy: $\delta<1$. Further, the exchange interaction is frustrated along the $c$-axis because  $J_2>0$, as confirmed by the observed period of the spiral ground state. The ratio of $J_1$ and $J_2$ is then fixed through the component of the magnetic propagation vector along the $c$-axis: $\cos(2\pi k_{z})=-J_1/4J_2$~\cite{Blundell}, yielding $J_2/J_1$~=~0.809.

The in-plane and out-of-plane magnetic susceptibility for the spin Hamiltonian \eqref{hamil} are  (see supplementary information),
\begin{subequations}\label{eq:susceptibility}
\begin{eqnarray}
 \chi _{{\nu \nu}}(\mathbf{q})  &=&-\frac{S^2}{4\hbar^2}\Big[\frac{\delta  J (\mathbf{q}-\mathbf{k})-J(\mathbf{k})}{(\omega +i\gamma)^2-\omega _0^2 (\mathbf{q}-\mathbf{k})}+\nonumber\\
  &+&\frac{\delta  J (\mathbf{q}+\mathbf{k})-J(\mathbf{k})}{(\omega +i\gamma )^2-\omega _0^2 (\mathbf{q}+\mathbf{k})}\Big],\\
\chi _{{zz}}({\mathbf{q}}) &=& \frac{ S^2 [-A({\mathbf{q}}) + J({\mathbf{k}})]}{{{\hbar ^2}[{{(\omega  + {{i}}\gamma )}^2} - \omega _0^2({\mathbf{q}})]}},
\end{eqnarray}
\end{subequations}
with $\nu=x,y$ and
\begin{equation}
\hbar {\omega _0}({\mathbf{q}}) = S \sqrt {[ - J({\mathbf{k}}) + \delta J({\mathbf{q}})]\left[ - J({\mathbf{k}}) + A(\mathbf{q})\right]},\label{eq:dispersion}
\end{equation}
where $A(\mathbf{q})=\frac{{J({\bf{q}} + {\bf{k}}) + J({\bf{q}} - {\bf{k}})}}{2}$. Here we have introduced a phenomenological damping constant $\gamma$ to account for the damping of spin wave by magnon-electron and magnon-magnon interactions. The spin wave dispersion is obtained from the poles of the magnetic susceptibility.

Finally, due to the small size of the magnetic Brillouin zone along the $c$-direction, Umklapp scattering occurs at the zone boundary, resulting in two additional spin wave branches $\omega_0(\mathbf{q}~ +~ \mathbf{k})$, and $\omega_0(\mathbf{q}~ -~ \mathbf{k})$, respectively. In principle, one expects additional branches at $\omega_0(\mathbf{q}\pm m \mathbf{k})$ with $m=0,1,2,3,...$, however, higher order Umklapp scattering is suppressed by the translational symmetry of the incommensurate magnetic spiral~\cite{Janoschek10}: any propagation of the helix by multiples of $\mathbf{k}$ can be compensated by changing the phase of the spiral by $2\pi$. This is consistent with our data, as we indeed observe three distinct spin wave branches~(Figs.~\ref{fig:spinwave} and~\ref{fig:cuts}).

The measured spin wave dispersion (Eq.~\ref{eq:dispersion}) can be well reproduced employing the following parameters $J_0=0.74$ meV, $J_1=0.1$ meV and $\delta=0.82$. Introducing a small, but non-resolution limited spin wave damping of $\hbar\gamma=0.15$ meV, additionally allows us to explain the full spin wave spectrum of \CeRh\ (Eq.~\ref{eq:susceptibility}) as demonstrated in Fig.~\ref{fig:spinwave}. We note, that a small discrepancy exists between this model and our data, which is the observed spin wave gap $\Delta$, that we have accounted for by shifting the calculated dispersion by $\Delta$. A gap is in principle not expected for an incommensurate spin spiral, which should always exhibit a gapless phason mode (Goldstone mode) due to the translational invariance~\cite{Janoschek10}. However, the presence of the gap is indeed consistent with specific heat measurements~\cite{Cornelius01}. The existence of a gap $\Delta$ due to defects is discussed in the supplementary material.

We will now discuss the implications of our results for understanding the phase diagram of \CeRh. A mean-field calculation based on the RKKY exchange interaction determined in the experiments here yields an antiferromagnetic ordering temperature $T^{\rm MF}_N$=$ J({\bf k}) S(S+1)/3 \simeq$ 9.1 K. Considering that  thermal fluctuations~\cite{Bao02}  tend to suppress the mean field ordering temperature by a factor of two  for a quasi two-dimensional system like \CeRh ($J_1/J_0 \sim 0.13$)~\cite{Pinaki03}, this is consistent  with the experimental value of $T_N$~=~3.8 K. We can compare this energy with the single-ion Kondo temperature $T_K$~=~0.15~K that has been determined from Ce$_{1-x}$La$_x$RhIn$_5$ with dilute amounts of Ce~\cite{Yang08}. Thus, at ambient pressure, the RKKY interaction is the dominant energy scale, and \CeRh\ exhibits well-localized $f$-electrons with local magnetic moments. This observation is consistent with de Haas-van Alphen measurements~\cite{Shishido05} that describe a small Fermi surface that does not include $f$-electrons. However, $T_K$ estimated via NMR measurements is 5~K~\cite{Curro03}, and the real value of the Kondo scale is likely in between those extremes. Further high-resolution neutron spectroscopy measurements above $T_N$ will be required to determine $T_K$ unambiguously.

It is also useful to compare our spin wave data on \CeRh\ with the related compounds in the family Ce$_m$$T_n$In$_{3m+2n}$. Measurements on the cubic material CeIn$_3$ ($m$~=~1, $n$~=~0) reveal a NN exchange interaction $J_0$~=~0.35 meV along the cubic diagonals~\cite{Knafo03}. The increased superconducting critical temperature $T_c$ in \CeRh\ ($m$~=~1, $n$~=~1)~ compared with its cubic building block CeIn$_3$ ($T_c$~=~0.2~K under pressure) is believed to be due to the reduced dimensionality that occurs because the CeIn$_3$ conducting layers are separated by an $T$In$_2$ layer in the Ce$T$In$_5$ family. Here we find that this reduced dimensionality is also reflected in the magnetic interactions, as the NN interaction $J_1$ along the $c$-axis is almost 10-fold reduced compared to the NN interaction $J_0$ within the tetragonal basal plane. In contrast, this trend does not continue for CePt$_2$In$_7$ ($m$~=~1, $n$~=~2), which exhibits a $T_c$~=~2.1 K (under pressure)~\cite{Sidorov13}. Thus, it would be useful to determine the exchange interactions for this compound, and verify if they exhibit less of a two-dimensional character. We note, that for HF systems the magnitude of the effective exchange constants can not easily be compared between two compounds, because they are sensitive towards the degree of $f$-electron delocalization controlled by the Kondo interaction.

While our results were obtained at ambient pressure they also allow us some insight into the unconventional superconducting state of \CeRh\ under pressure. In the case that bosonic degrees of freedom such as the spin waves observed here would be involved in mediating the superconductivity in \CeRh, it is expected that the coupling between the electronic quasiparticles and the spin waves is retarded~\cite{Steglich13}. The Fermi velocity of \CeRh\ in the superconducting state under pressure was estimated as $v_F$~$\approx$~39 meV\AA~\cite{Park08a}. We can use the measured spin wave dispersion to estimate the spin velocity in the tetragonal basal plane as $v_{\rm SW}$~=~8(1) meV\AA, demonstrating that the magnetic excitations are indeed retarded with respect to the electronic quasiparticles. Notably, the retardation is of similar magnitude as the values identified for CeCu$_2$Si$_2$, for which the superconductivity is believed to be mediated by magnetic fluctuations~\cite{Stockert11}, suggesting that the the spin excitations in \CeRh\ are a possible candidate for a superconducting glue.

Finally, we note that the frustrated exchange interactions identified in \CeRh\ will allow us to study new interesting physics such as micro-phases and competing magnetic structures and their interplay with strongly correlated electron phenomena such as unconventional superconductivity~\cite{Park12}. They further pave the way to quantitatively investigate the recently proposed model for HF materials that attempts to capture the physics of local and conventional quantum criticality in a global zero-temperature phase diagram~\cite{Coleman10, Si14}. Notably, the theory suggest that the global phase diagram may be controlled via two distinct tuning parameters, where the first one is determined via the competition between RKKY and Kondo interactions, and the latter is the amount of magnetic frustration. Careful inelastic neutron scattering measurements in \CeRh\ as function of various tuning parameters such as pressure, magnetic field or chemical substitution will allow the characterisation of the amount of frustration throughout the phase diagram.

In conclusion, we have quantitatively determined the magnitude of the magnetic exchange interactions in \CeRh, which is a member of the prototypical family Ce$T$In$_5$ of HF compounds. This first microscopic measurement of the magnetic exchange interactions in \CeRh\ is an important step towards a quantitative understanding of the complex phase diagrams observed in the Ce$T$In$_5$ family and related compounds, and will be a key parameter for testing the validity of current and future theories.

\begin{acknowledgments}
Work at Los Alamos National Laboratory (LANL) was performed under the auspices of the U.S. DOE, OBES, Division of Materials Sciences and Engineering and funded in part by the LANL Directed Research and Development program. Research conducted at SNS (CNCS instrument) was sponsored by the Scientific User Facilities Division, Office of Basic Energy
Sciences, US Department of Energy. We are grateful to Monika Hartl and Luke Daemon for technical support when co-aligning the sample mosaic by means of x-ray diffraction. We further acknowledge useful discussions with Jon Lawrence.
\end{acknowledgments}

%\bibliographystyle{prsty}

%\putbib[maintext]
\end{bibunit}

\begin{bibunit}
\balancecolsandclearpage
\section*{SUPPLEMENTARY MATERIAL}

\setcounter{section}{0}

We describe details of the $J_1$-$J_2$ model used to fit the spin waves in \CeRh and supplement analysis that support our discussion of the measured spin wave spectrum and magnetic exchange interaction in \CeRh.

\section{Spin wave dispersion for a magnetic spiral structure}
In this section, we provide details of the spin wave calculation. The spiral structure is described by $S_x(\mathbf{r})=S\cos(2\pi\mathbf{k}\cdot\mathbf{r})$, $S_y(\mathbf{r})=S\sin(2\pi\mathbf{k}\cdot\mathbf{r})$ and $S_z(\mathbf{r})=0$, with the ordering wavevector $\mathbf{k}=(0.5,\ 0.5,\ 0.297)$ and $S=1/2$. Here $\mathbf{r}$ is renormalized by the lattice constant along each crystallographic axis. We introduce a local reference frame such that the local quantization axis ($z$-axis) is parallel to the magnetic spiral and the local $x$ axis remains in the $x$-$y$ plane of the original frame. The spin components in the two reference frames are related by the following expressions:
\begin{equation}\label{seq1}
{S_x} =  - {\tilde{S}_x}\sin(2\pi \mathbf{k}\cdot\mathbf{r}) + {\tilde{S}_z}\cos(2\pi \mathbf{k}\cdot\mathbf{r}),
\end{equation}
\begin{equation}\label{seq2}
{S_y} = {\tilde{S}_x}\cos(2\pi \mathbf{k}\cdot\mathbf{r}) + {\tilde{S}_z}\sin(2\pi \mathbf{k}\cdot\mathbf{r}),
\end{equation}
\begin{equation}\label{seq3}
{S_z} = {\tilde{S}_y},
\end{equation}
where $\tilde{\mathbf{S}}$ is the spin operator in the local reference frame. The Hamiltonian in Eq. (1) of the main text can be expressed as
\begin{align}\label{seq4}
{\mathcal H} = \sum_{i<j} {J_{ij}}(\tilde{S}_{x, i}\tilde{S}_{x, j} + \tilde{S}_{z, i}\tilde{S}_{z, j})\cos [2\pi \mathbf{k}(\mathbf{r}_i-\mathbf{r}_j)] \nonumber\\
+\sum_{i<j} {J_{ij}} (\tilde{S}_{z,i}\tilde{S}_{x, j} - \tilde{S}_{x, i}\tilde{S}_{z,j})\sin [2\pi \mathbf{k}(\mathbf{r}_i-\mathbf{r}_j)] \nonumber
\\
+\sum_{i<j}  \delta {J_{ij}}\tilde{S}_{y, i}\tilde{S}_{y, j}.
\end{align}
To obtain the spin wave spectrum, we employ the Holstein-Primakoff transformation
\begin{equation}\label{seq5}
\tilde{S}_i^ -  = \sqrt {2S} a_i^\dag,\ \tilde{S}_i^+  = \sqrt {2S} {a_i},\ \tilde{S}_i^z = S - a_i^\dag {a_i},
\end{equation}
with
\begin{equation}\label{seq6}
\tilde{S}_i^ \pm  = \tilde{S}_{x,i} \pm {i}\tilde{S}_{y, i}.
\end{equation}
By expanding up to quadratic order in the  Holstein-Primakoff bosons ($1/S$ expansion), we can approximate $\mathcal{H}$ by $\mathcal{H}\simeq\mathcal{H}_0+\mathcal{H}_2$, with
\begin{equation}\label{seq7}
\mathcal{H}_0= - \frac{N}{2}{S^2}J(\mathbf{k}),
\end{equation}
\begin{equation}\label{seq8}
\mathcal{H}_2=\frac{S}{2}\mathop \sum \limits_q \left( {\begin{array}{*{20}{c}}
	{{a_{ - q}}}&{a_q^\dag }
	\end{array}} \right)\left( {\begin{array}{*{20}{c}}
	A_q&B_q\\
	B_q&A_q
	\end{array}} \right)\left( {\begin{array}{*{20}{c}}
	{a_{ - q}^\dag }\\
	{{a_q}}
	\end{array}} \right),
\end{equation}
where $N$ is the number of spins and
\begin{align}\label{seq9}
J(\mathbf{q})=2J_0 [\cos(2\pi q_x)+\cos(2\pi q_y)]\nonumber\\
+2J_1\cos(2\pi q_z)+2J_2\cos(4\pi q_z),
\end{align}
\begin{equation}\label{seq10}
{A_q} = \frac{{[J(\mathbf{q} - \mathbf{k}) + J(\mathbf{q} + \mathbf{k})]}}{4} + \frac{{\delta J(\mathbf{q})}}{2} - J(\mathbf{k}),
\end{equation}
\begin{equation}\label{seq11}
B_q=\frac{{[J(\mathbf{q} - \mathbf{k}) + J(\mathbf{q} + \mathbf{k})]}}{4} - \frac{{\delta J(\mathbf{q})}}{2}.
\end{equation}
The interactions $J_0$, $J_1$ and $J_2$ are defined in the main text.  Eq. \eqref{seq8} is diagonalized by applying a standard Bogoliubov transformation:
\begin{equation}\label{seq12}
{a_q} = {u_q}{b_q} - {v_q}b_{ - q}^{\dag }\ \text{and}\ \ a_q^\dag  = {u_q}b_q^\dag  - {v_q}{b_{ - q}},
\end{equation}
with $u_q=\cosh\theta_q$, $v_q=\sinh\theta_q$ and $\tanh(2\theta_q)=B_q/A_q$.  The diagonal form of  $\mathcal{H}_2$ is
\begin{equation}\label{seq13}
\mathcal{H}_2=\hbar \sum_q \omega_0(\mathbf{q}) \left(b_q^\dag {b_q} + \frac{1}{2}\right).
\end{equation}
with a  dispersion in the spin wave (in the local reference frame) given by
\begin{equation}\label{seq14}
\hbar \omega_0(\mathbf{q})= S \sqrt {(A_q+B_q)(A_q-B_q)}.
\end{equation}

To describe the neutron scattering measurements, we calculate the magnetic susceptibility in the original reference frame. By using Eqs. \eqref{seq1}, \eqref{seq2} and \eqref{seq3}, the susceptibility components, , $\chi_{xx}$, $\chi_{yy}$, $\chi_{zz}$, in the original reference frame can be expressed in terms of the components, $\tilde{\chi}_{xx}$, $\tilde{\chi}_{zz}$, in the local reference frame:
\begin{align}\label{seq15}
\chi _{xx}(\mathbf{q},\omega ) = \chi _{yy}(\mathbf{q},\omega ) \equiv \langle {S_x}(\mathbf{q},\omega ){S_x}( - \mathbf{q}, - \omega )\rangle  \nonumber\\
= \frac{1}{4}\left[\tilde{\chi} _{xx}(\mathbf{q} - \mathbf{k},\omega ) + \tilde{\chi} _{xx}(\mathbf{q} + \mathbf{k},\omega )\right],
\end{align}
\begin{equation}\label{seq16}
\chi _{zz}(\mathbf{q},\omega ) = \tilde{\chi} _{yy}(\mathbf{q},\omega ).
\end{equation}
$\tilde{\chi} _{xx}$ and $\tilde{\chi} _{yy}$ are straightforwardly obtained from the diagonal form of  the Hamiltonian in Eq. \eqref{seq13}
\begin{equation}\label{seq17}
\tilde{\chi} _{xx}(\mathbf{q},\omega ) = -\frac{S^2}{\hbar^2}\frac{\delta  J (\mathbf{q})-J(\mathbf{k})}{(\omega +i\gamma)^2-\omega _0^2 (\mathbf{q})},
\end{equation}
\begin{equation}\label{seq18}
\tilde{\chi} _{yy}(\mathbf{q},\omega ) = \frac{S^2}{\hbar^2}\frac{{  J({\bf{k}}) - 0.5[J({\bf{q}} + {\bf{k}}) + J({\bf{q}} - {\bf{k}})] }}{{{{(\omega  + {{i}}\gamma )}^2} - \omega _0^2({\bf{q}})}}.
\end{equation}
Here we have introduced an phenomenological parameter $\gamma$ to account for the finite damping  due to magnon-magnon,  magnon-electron, and magnon-lattice interactions. The spin wave dispersion is obtained from  the poles of $\chi_{xx}$ and $\chi_{zz}$. There are three branches: $\omega_0(\mathbf{q})$ and $\omega_0(\mathbf{q}\pm \mathbf{k})$. The neutron cross-section is given by \cite{JensenBook}
\begin{equation}\label{seq19}
\frac{{{{d}^2}\sigma }}{{{d}E{d}\Omega }} \propto \sum_{\alpha,\beta}(\delta_{\alpha\beta} - \hat{q}_\alpha\hat{q}_\beta)\mathrm{Im}\left[{\chi _{\alpha\beta}}({\bf{q}},\omega)\right],
\end{equation}
where $\hat{\mathbf{q}}=\mathbf{q}/|\mathbf{q}|$.

\section{Possible origin of the spin wave gap}
An incommensurate magnetic spiral structure leads to a gapless phason mode due to the spontaneous breaking of a continuous symmetry. One possible mechanism for explaining the spin gap observed in the neutron scattering experiments is the presence of defects that explicitly break the translational invariance along the $c$-axis. These defects generate a  pinning potential for the magnetic spiral and this pinning gives rise to a gap in the spin wave spectrum. Here we consider the pinning of a magnetic spiral in three dimensions. The effective Lagrangian density for the phason mode $\phi$ in the presence of pinning can be written as
\begin{equation}
\mathcal{L}=\mathcal{N}\left[\frac{{{\hbar ^2}}}{2}{({\partial _t}\phi )^2} - \frac{{{\hbar ^2}v_s^2}}{2}{(\nabla \phi )^2}\right] - {V_{\text{pin}}}(\phi ),
\end{equation}
where $v_s$ is the spin wave velocity in the long wavelength limit and $\mathcal{N}$ is the magnon density of states. We estimate $\mathcal{N}\sim 1/(J_0 a^2 c)$ with $a$ and $c$ the lattice parameters. The random pinning potential can be modelled as
\begin{equation}
{V_{\text{pin}}}(\phi ) = \mathop \sum \limits_i {V_{p0}}f(\phi )\delta ({\bf{r}} - {{\bf{r}}_i}),
\end{equation}
where $V_{p0}$ is the pinning strength and $f(\phi)$ is some function of the order of unity, which depends on the particular microscopic pinning mechanism. Here we have considered uncorrelated defects. If $\phi$ adjusts to all the pinning potentials, the energy density gain is ${\mathcal{E}_{\text{pin}}} \approx   {V_{p0}}{\rho _d}$, where $\rho_d$ is the density of pinning centers. The energy density cost due to the distortion is ${\mathcal{E}_{\text{kin}}} \approx \mathcal{N}{\hbar ^2}v_s^2\rho _d^{2/3}$. Depending on the ratio between these two energy scales, the pinning can be  strong,
\begin{equation}
\eta  \equiv \frac{{{\mathcal{E}_{\text{pin}}}}}{{{\mathcal{E}_{\text{kin}}}}} = \frac{{{V_{p0}}\rho _d^{1/3}}}{{\mathcal{N}{\hbar ^2}v_s^2}} >  > 1,
\end{equation}
or  weak, $\eta <<1$.

In the strong pinning regime, the spiral adjusts locally to the pinning potential and the total pinning energy is
\begin{equation}
{\mathcal{E}_{p}} \approx   {V_{p0}}{\rho _d}.
\end{equation}
The spiral is then distorted over a length scale ${L_0} \approx \rho _d^{ - 1/3}$, which becomes the correlation length for the spin-spin correlation function.

In the weak pinning regime, $L_0$ can be estimated by using a scaling argument. The energy gain in the volume $L_0^3$ due to the random pinning potential is ${\mathcal{E}_{\text{pin}}} \approx   {{{V_{p0}}{{(L_0^3{\rho _d})}^{1/2}}}}{{L_0^{-3}}}$, while the energy cost due to the distortion is ${\mathcal{E}_{\text{kin}}} \approx {{\mathcal{N}{\hbar ^2}v_s^2}}{{L_0^{-2}}}$. The optimal distortion arises from the best compromise between $\mathcal{E}_{\text{pin}}$ and $\mathcal{E}_{\text{kin}}$, yielding ${L_0} = {{16 {{\mathcal{N}^2}{\hbar ^4}v_s^4}}}/{{9 {{\rho _d}V_{{\rm{p0}}}^2} }}$. The corresponding optimal pinning energy is
\begin{equation}
{{ \mathcal E}_p} =   \frac{{27}}{{256}}\frac{{V_{p0}^4\rho _d^2}}{{{\mathcal{N}^3}{\hbar ^6}v_s^6}}.
\end{equation}

Knowing the pinning energy, $\mathcal{E}_p$, we can express the Lagrangian for the phason mode as
\begin{equation}
{\cal L} = \mathcal{N}\left[\frac{{{\hbar ^2}}}{2}{({\partial _t}\phi )^2} - \frac{{{\hbar ^2}v_s^2}}{2}{(\nabla \phi )^2}\right] -\frac{1}{2}{{ \mathcal E}_p}\phi^2,
\end{equation}
The resulting dispersion for the phason,
\begin{equation}
{\hbar ^2}{\omega ^2} = \frac{\mathcal{E}_p}{\mathcal{N}} + {\hbar ^2}v_s^2{q^2},
\end{equation}
has a gap of magnitude
\begin{equation}
\Delta = \sqrt{\frac{\mathcal{E}_p}{\mathcal{N}} }.
\end{equation}
$\Delta$ is not universal, as it depends on details of the random pinning potential, such as pinning density and pinning strength, which vary from sample to sample. To compare with our experiments, we need to know these parameters for a given sample. Such information is not available for the moment. However, we can anticipate that, if this is the mechanism for explaining the observed spin gap, the value of $\Delta$ should vary from one sample to another depending on the sample quality.

In the weak pinning regime, we have $\Delta L_0=\hbar v_s/\sqrt{3}$. By using the measured $\Delta=0.25$ meV and $\hbar v_s = 4.95$ meV$\AA$  along $k_z$, we estimate $L_0\sim 10\ \AA$. Because such  a high density of defects is incompatible with existing experiments \cite{Moshopoulou:01},  we exclude the weak-pinning regime as a possible scenario for explaining the measured spin gap of CeRhIn$_5$. In the strong pinning regime, the gap is given by
\begin{equation}
\Delta=\sqrt{V_{p0}\rho_d/\mathcal{N}}.
\end{equation}
If we take $\rho_d\sim 0.01/(a^2 c)$, \cite{Moshopoulou:01} we can estimate $V_{p0}\approx 10 J_0\approx 7\ \mathrm{meV}$.
In other words, the strong-pinning regime could explain the experimentally observed energy gap in the spin wave spectrum if the strength of the pining potential is an order of magnitude larger than the dominant exchange constant.

\end{bibunit}


\begin{thebibliography}{10}

\bibitem{Loehneysen07}
  H. v. L{\"o}hneysen, A. Rosch, M. Vojta and P. W{\"o}lfle,
  Rev. Mod. Phys. {\bf 79}, 1015 (2007).

\bibitem{Gegenwart08}
  P. Gegenwart, Q. Si and F. Steglich,
  Nat. Phys. {\bf 4}, 892 (2008).

\bibitem{Doniach77}
  S. Doniach,
  Physica B {\bf 91}, 231 (1977).

\bibitem{Kondo64}
  J. Kondo,
  Prog. Theor. Phys. {\bf 32}, 37 (1964).

\bibitem{Fisk86}
  Z. Fisk, H. R. Ott, T. M. Rice and J. L. Smith,
  Nature {\bf 320}, 124 (1986).

\bibitem{Mathur98}
  N. D. Mathur, F. M. Grosche, S. R. Julian, I. R. Walker, D. M. Freye, R. K. W. Haselwimmer, and G. G. Lonzarich,
  Nature {\bf 394},  39  (1998).

\bibitem{Pfleiderer09}
  C. Pfleiderer,
  Rev. Mod. Phys. {\bf 81}, 1551 (2009).

\bibitem{Dahm09}
  T. Dahm, V. Hinkov, S. V. Borisenko, A. A. Kordyuk, V. B. Zabolotnyy, J. Fink, B. Büchner, D. J. Scalapino, W. Hanke, and B. Keimer,
  Nat. Phys. {\bf 5}, 217 (2009).

\bibitem{Wang13}
  M. Wang {\em et al.},
  Nat. Commun. {\bf 4}, 3874 (2013).

\bibitem{Si01}
  Q. Si, S. Rabello, K. Ingersent, and J. Lleweilun Smith,
  Nature {\bf 413}, 804 (2001).

\bibitem{Coleman10}
  P. Coleman and A. H. Nevidomskyy,
  J. Low Temp. Phys. {\bf 161}, 182 (2010).

\bibitem{Si14}
  Q. Si, J. H. Pixley, E. Nica, S. J. Yamamoto, P. Goswami, R. Yu, and S. Kirchner,
  J. Phys. Soc. Jpn. {\bf 83}, 061005 (2014).

\bibitem{Broholm87}
  C. Broholm, J. K. Kjems, G. Aeppli, Z. Fisk, J. L. Smith, S. M. Shapiro, G. Shirane, and H. R. Ott,
  Phys. Rev. Lett. {\bf 58}, 917 (1987).

\bibitem{Van Dijk00}
  N.H. van Dijk, B. F{\aa}k, T. Charvolin, P. Lejay, J.M. Mignot, and B. Hennion,
  Physica B {\bf 241}, 808 (1998).

\bibitem{Knafo03}
  W. Knafo, S. Raymond, B. F{\aa}k, G. Lapertot, P. C. Canfield and J. Flouquet,
  J. Phys.: Condens. Matter {\bf 15}, 3741 (2003).

\bibitem{Fak08}
  B. F{\aa}k, S. Raymond, D. Braithwaite, G. Lapertot, and J.-M. Mignot,
  Phys. Rev. B {\bf 78}, 184518 (2008).

\bibitem{Stockert11}
  O. Stockert {\em et al.},
  Nat. Phys. {\bf 7}, 119–124 (2011).

\bibitem{Thompson:12}
  J. D. Thompson and Z. Fisk,
  J. Phys. Soc. Jpn. {\bf 81}, 011002 (2012).

\bibitem{Bao00}
  W. Bao, P. G. Pagliuso, J. L. Sarrao, J. D. Thompson, Z. Fisk, J. W. Lynn, and R. W. Erwin,
  Phys. Rev. B {\bf 62}, R14621(R) (2000).

\bibitem{Park06}
  T. Park, F. Ronning, H. Q. Yuan, M. B. Salamon, R. Movshovich, J. L. Sarrao and J. D. Thompson,
  Nature {\bf 440}, 65 (2006).

\bibitem{Park08}
  T. Park, V. A. Sidorov, F. Ronning, J.-X. Zhu, Y. Tokiwa, H. Lee, E. D. Bauer, R. Movshovich, J. L. Sarrao, and J. D. Thompson,
  Nature {\bf 456}, 366 (2008).

\bibitem{Stock08}
  C. Stock, C. Broholm, J. Hudis, H. J. Kang, and C. Petrovic,
  Phys. Rev. Lett. {\bf 100}, 087001 (2008).

\bibitem{Shishido05}
  H. Shishido, R. Settai, H. Harima, and Y. $\bar{\mathrm{O}}$nuki,
  J. Phys. Soc. Jpn. {\bf 74}, 1103 (2005).

\bibitem{Ehlers11}
  G. Ehlers, A. A. Podlesnyak, J. L. Niedziela, E. B. Iverson, and P. E. Sokol,
  Review of Scientific Instruments {\bf 82}, 085108 (2011).

\bibitem{Bao02}
  W. Bao, G. Aeppli, J. W. Lynn, P. G. Pagliuso, J. L. Sarrao, M. F. Hundley, J. D. Thompson, and Z. Fisk,
  Phys. Rev. B {\bf 65}, 100505(R) (2002).

\bibitem{Willers10}
  T. Willers {\em et al.},
  Phys. Rev. B {\bf 81}, 195114 (2010).

\bibitem{Blundell}
  S. Blundell, ``Magnetism in Condensed Matter'', page 99, Oxford University Press  (2001).

\bibitem{Janoschek10}
  M. Janoschek, F. Bernlochner, S. Dunsiger, C. Pfleiderer, P. B{\"o}ni, B. Roessli, P. Link, and A. Rosch,
  Phys. Rev. B {\bf 81}, 214436 (2010).

\bibitem{Cornelius01}
  A. L. Cornelius, P. G. Pagliuso, M. F. Hundley, and J. L. Sarrao,
  Phys. Rev. B {\bf 64}, 144411 (2001).

\bibitem{Pinaki03}
  P. Sengupta, A. W. Sandvik, and R. R. P. Singh, Phys. Rev. B {\bf 68}, 094423 (2003).

\bibitem{Yang08}
  Yi-feng Yang, Z. Fisk, H.-O. Lee, J. D. Thompson and D. Pines,
  Nature {\bf 454}, 611 (2008).

\bibitem{Curro03}
  N. J. Curro, J. L. Sarrao, J. D. Thompson, P. G. Pagliuso, \v{S}. Kos, Ar. Abanov, and D. Pines,
  Phys. Rev. Lett. {\bf 90}, 227202 (2003)

\bibitem{Sidorov13}
  V. A. Sidorov, Xin Lu, T. Park, Hanoh Lee, P. H. Tobash, R. E. Baumbach, F. Ronning, E. D. Bauer, and J. D. Thompson,
  Phys. Rev. B {\bf 88}, R020503 (2013).

\bibitem{Steglich13}
  F. Steglich, O. Stockert, S. Wirth, C. Geibel, H. Q. Yuan, S. Kirchner, and Q. Si,
  J. Phys: Conf. Series {\bf 449}, 012028 (2013).

\bibitem{Park08a}
  T. Park, M. J. Graf, L. Boulaevskii, J. L. Sarrao, and J. D. Thompson,
  PNAS {\bf 105}, 6825 (2008).

\bibitem{Park12}
  T. Park, H. Lee, I. Martin, X. Lu, V. A. Sidorov, K. Gofryk, F. Ronning, E. D. Bauer, and J. D. Thompson,
  Phys. Rev. Lett. {\bf 108}, 077003 (2012).


\end{thebibliography}

\begin{thebibliography}{10}

\bibitem{JensenBook}
J. Jensen and A. R. Mackintosh, {\it Rare Earth Magnetism}, Oxford Science Publ., New York (1991).

\bibitem{Moshopoulou:01}
  E. G. Moshopoulou, Z. Fisk, J. L. Sarrao, and J. D. Thompson,
  J. Solid State Chem. {\bf 158}, 25 (2001).

\end{thebibliography}
\end{document}